\documentstyle[epsf,12pt]{article}
\begin{document}

\begin{flushright}
UT-Komaba 98- 23   \\
ICRR-Report-452-99-10
\end{flushright}
\vskip 0.1in
\begin{center} 
{\Large{\bf Localized and Extended States in One-Dimensional 
Disordered System: }}\\
\vskip 0.15cm
{\Large{\bf Random-Mass Dirac Fermions }}
\vskip 1.5cm

{\large Koujin Takeda}\footnote{e-mail address:takeda@ctpc1.icrr.u-tokyo.ac.jp} \\
Institute for Cosmic Ray Research, University of Tokyo, Tanashi, 
Tokyo, 188-8502 Japan\\
{\large Toyohiro Tsurumaru}\footnote{e-mail address:tsuru@tanashi.kek.jp} \\
Institute of Particle and Nuclear Studies, High Energy Accelerator Research
Organization (KEK), Tanashi Branch, Tanashi,
Tokyo, 188-8501 Japan\\
{\large Ikuo Ichinose}\footnote{e-mail address:ikuo@hep1.c.u-tokyo.ac.jp}\\
Institute of Physics, University of Tokyo, 
Komaba, Tokyo, 153-8902 Japan\\
{\large Masaomi Kimura}\footnote{e-mail address:
masaomi@ctpc1.icrr.u-tokyo.ac.jp} \\
Institute for Cosmic Ray Research, University of Tokyo, Tanashi, 
 Tokyo, 188-8502 Japan
 \\

\end{center}

\newpage

\begin{center} 
\begin{bf}
Abstract
\end{bf}
\end{center}
System of Dirac fermions with random-varying mass is studied in detail.
We reformulate the system by transfer-matrix formalism.
Eigenvalues and wave functions are obtained numerically for various 
configurations
of random telegraphic mass $m(x)$.
Localized and extended states are identified.
For quasi-periodic $m(x)$, low-energy wave functions are also 
quasi-periodic and extended,
though we are {\em not} imposing the periodic boundary condition 
on wave function.
On increasing the randomness of the varying mass, states lose periodicity 
and most of them
tend to localize.
At the band center or the low-energy limit, there exist extended states which 
have more than one peak spatially separate with each other 
comparatively large distance.
Numerical calculations of the density of states 
and ensemble averaged Green's functions are explicitly given.
They are in good agreement with
analytical calculations by using the supersymmetric methods
and exact form of the zero-energy wave functions.

\setcounter{footnote}{0}
\setcounter{equation}{0}
\section{Introduction}

Random disordered system is one of the most important 
problem in condensed matter physics.
In two or lower dimensions, it is expected that almost all states are 
localized by the existence of random potentials, 
and study of these random systems obviously requires
non-perturbative methods.
In most of cases there exists no controllable parameter for analytical 
calculation.
However recently, exact results have been obtained in certain 
interesting disordered 
systems in low dimensions\cite{Ludwig,Comtet,BF}.
There almost all states are localized as expected from the 
general consideration,
and extended states exist only at isolated points in 
physical parameter region. 
Mathematical tools like supersymmetric (SUSY) methods {\it etc.} play an 
important role
in averaging quenched disorders.

The random hopping tight-binding (RHTB) model in one dimension 
is one of the most extensively studied model\cite{BF,Shelton,Mathur}.
This model is related with a random spin chain, a doped spin ladder, 
{\it etc.} \cite{Shelton,Gog}.
By using SUSY methods, the density of states (DOS)
and single-fermion Green's
functions were obtained for white-noise random hopping\cite{BF}.
Recently, the case of non-locally correlated disorders was studied
by extending SUSY methods\cite{IK,IK2}.
This generalization is practically important because in almost all real
systems random disorders have nonlocal correlations.

In this paper we shall revisit the system of random-mass Dirac fermion,
which describes low-energy excitations near the band center 
of the RHTB model.
However our approach in this paper is somewhat 
different from those given so far,
and we hope that our study and previous works are  
complementary with each other.
In most of studies, average over quenched disorder is taken
in order to obtain physical quantities.
In this paper however, we shall first study quantum mechanical states in 
various {\em fixed}
random disorder backgrounds. 
To end this, we use the transfer matrix (TM) formalism. 
The TM formalism was used to calculate the transmission 
and reflection coefficients of current incident to a 
one-dimensional disordered 
system having a certain momentum\cite{erdoes}. 
In this paper, we shall extend the TM formalism and apply it  to
bound states and/or localized states to find their energy eigenvalues
and wave functions.
On controlling randomness, we can see how states tend to localize
and what kind of states remain extended and contribute to long-distance
correlations.
To this end, results of our previous study on non-locally correlated 
disorder case
is quite useful.
After this observation, we shall take an ensemble average with 
respect to random variables and calculate the DOS
and Green's functions.
We show that numerical calculations are in good agreement with
analytical results in Refs.\cite{IK,BF} which were obtained
by the SUSY methods and explicit form of zero-energy wave functions.
This verifies that the SUSY methods give correct results. 

This paper is organized as follows.
In Sect.2, field equations of random-mass Dirac fermion is given.
In Sect.3, we shall consider field equations in telegraphic configurations of 
random mass and reformulate the system by introducing the 
TM formalism. 
In the TM formalism, the eigenvalue problem of the Dirac field
reduces to a simple matrix equation.
Wave functions are also easily obtained by numerical calculation
in the TM formalism. 
Results of numerical calculation are given in Sect.4.
We identify localized and extended states from the wave functions.
In order to see this, our previous studies on non-locally correlated 
disorder are quite useful.
Especially it is shown that the DOS obtained by the
numerical methods is in good agreement with the previous analytical calculation.We also obtain the ensemble averaged Green's functions at vanishing
energy or for nodeless states and compare them with analytic
expression in Ref.\cite{BF}.
They are in good agreement though the exact zero-energy wave functions
used in the analytic calculation are {\em not} always normalizable
contrary to those in the numerical calculation.
Section 5 is devoted for discussion.
We shall discuss relationship between the localization length and the number
of states with energy below $E$.
Some application of the results for spin chains is also discussed.

\setcounter{equation}{0}
\section{Field equations}
We shall consider a Dirac fermion in one spatial dimension with 
coordinate-dependent mass whose Hamiltonian is given by,
\begin{eqnarray}
{\cal H}&=&\int dx \psi^\dagger h\psi,\\
h&=&-i\sigma^z\partial_x+m(x)\sigma^y,
\end{eqnarray}
where $\vec{\sigma}$ are the Pauli matrices. 
Defining each component of $\psi$ as $\psi=(u,v)$, we can write the 
 Dirac equation as,
\begin{eqnarray}
\left(\ \frac{d}{dx}\ +\ m(x)\ \right)\ u(x)&=&Ev(x),\nonumber\\
\left(\ -\frac{d}{dx}\ +\ m(x)\ \right)\ v(x)&=&Eu(x). \label{eq:dirac1}
\end{eqnarray}
From Eqs.(\ref{eq:dirac1}), we obtain 
Schr$\ddot{\mbox{o}}$dinger equations,
\begin{eqnarray}
\left(\ -\frac{d^2}{dx^2}\ -m'(x)+m^2(x)\ \right)\ u(x)&=&E^2u(x),\nonumber\\
\left(\ -\frac{d^2}{dx^2}\ +m'(x)+m^2(x)\ \right)\ v(x)&=&E^2v(x),
\label{eq:schroedinger1}
\end{eqnarray}
where prime denotes derivative with respect to the space coordinate. 
The exact solutions to (\ref{eq:dirac1}) for $E=0$ are easily found
though they are not always normalizable,
\begin{equation}
u,v\propto\exp\left[\ \pm\int dx\ m(x)\ \right]. \label{eq:exactsol}
\end{equation}
However, solutions for positive eigen-values of $E$ cannot be represented in 
a simple form as in (\ref{eq:exactsol}).

On the other hand, when $m(x)$ has a soliton-like configuration, 
such as\footnote{This configuration of $m(x)$ is a soliton solution in the  
$\lambda\phi^4$
Higgs potential of double-well form\cite{NS}.}
\begin{equation}
m(x)=\bar{m}\tanh(\lambda \bar{m}x),\label{eq:soliton1}
\end{equation}
the lowest-energy eigenstate for $\lambda=1$ is a ``bound" state and 
explicitly given by
\begin{equation}
\label{eq:boundstate}
u(x)=\frac1{\cosh^2 \bar{m}x}, \;\;\; v(x)=0.
\end{equation}
We call the configuration (\ref{eq:soliton1}) for $\lambda>0$ anti-soliton 
and that for 
$\lambda<0$ soliton. 
It is one of our conjecture that localized states in the random potential,
$V_r(x)=\pm m'(x)+m^2(x)$, are essentially
given by linear combinations of the above bound states in soliton and 
anti-soliton configuration of $m(x)$. 
To see this, we approximate the random potential to be a repetition of 
anti-soliton-soliton 
like configurations (\ref{eq:soliton1}) and further, deform it to a series 
of the step functions by 
taking the limit $\lambda\to\infty$;
\begin{equation}
m(x)=\sum_i\bar{m}(\theta(x-\alpha_i)-1)+\sum_j\bar{m}
(\theta(-x+\beta_j)-1)+m_0,
\label{stepm}
\end{equation}
where
$\alpha_i$ and $\beta_j$ are positions of solitons 
and anti-solitons, respectively (see Fig.1). 
In the subsequent sections, we shall solve the
Schr$\ddot{\mbox{o}}$dinger equations 
in the random background (\ref{stepm}).
\begin{figure}
\label{fig:example}
\unitlength=1cm
\begin{picture}(15,5)
\unitlength=1mm
\centerline{
\epsfysize=5cm
\epsfbox{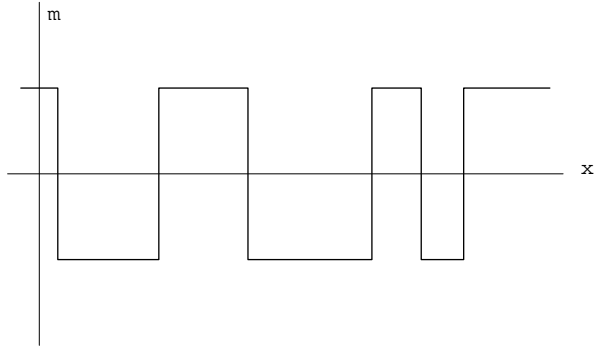}
}
\end{picture}
\caption{An example of configurations of solitons and 
anti-solitons, or ``rectangular barriers" (Eq.(\ref{stepm})).}
\end{figure}
\setcounter{equation}{0}
\section{Eigenfunctions}
For nonzero $E$, solutions of the Dirac equation (\ref{eq:dirac1}) 
cannot be 
expressed in a general form as in (\ref{eq:exactsol}). 
However, if we take the limit $\lambda\to\infty$, $m(x)$ takes alternatively 
two values 
$m_0+\bar{m}$ and $m_0-\bar{m}$ on intervals whose lengths are positive 
random variables. 
This drastically simplifies the problem and we can calculate the energy 
eigenfunctions for nonzero $E$ systematically using transfer matrices. 
In the first half of this section, solution in the background $m(x)$ 
with one step function 
is considered as a special case, and in the second half, 
it is shown that by connecting so 
obtained solutions, one can obtain solutions for arbitrary patterns of 
anti-soliton-soliton pairs.
\subsection{Step-functional background}
We study the Dirac equation (\ref{eq:dirac1}) with a mass 
configuration
which has a ``domain wall" at $x=0$,
\begin{equation}
m(x)=\bar{m}(2\theta(-x)-1)+m_0,\label{eq:stepfunct}
\end{equation}
where  $\theta(x)$ is the Heaviside function,
\begin{equation}
\theta(x)=\left\{
\begin{array}{ccc}
0&\ &(x<0)\\
1&\ &(0<x)
\end{array}
\right.,
\end{equation}
and it is assumed that
\[\bar{m}>m_0>0.\]
Substituting (\ref{eq:stepfunct}) into (\ref{eq:schroedinger1}), we obtain 
a Schr$\ddot{\mbox{o}}$dinger equation with a potential which is 
a combination of 
the delta function and the step function,
\begin{equation}
V(x)=\mp m'(x)+m^2(x)=\pm2\bar{m}\delta(x)+4m_0\bar{m}
\theta(-x)+(\bar{m}-m_0)^2.
\end{equation}
For such a field equation with a delta-function-type potential, 
the continuity of the wave function $u(x)$ leads automatically to 
the conditions 
on its derivative $u'(x)$. 
For example, if we rewrite the Schr$\ddot{\mbox{o}}$dinger equation as
\begin{equation}
\left(-\frac{d^2}{dx^2}+2\bar{m}\ \delta(x)+4m_0\bar{m}\ \theta(-x)\right)
\ u(x)=(E^2-(\bar{m}-m_0)^2)\ u(x)
\label{Seq}
\end{equation}
and integrate Eq.(\ref{Seq}) with respect to $x$ in the range
$-\epsilon\le x\le\epsilon$, we have
\begin{equation}
-u'(0+\epsilon)+u'(0-\epsilon)+2\bar{m}\ u(0)+4m_0\bar{m}\int^0_x dx\ u(x)
=(E^2-(\bar{m}-m_0)^2)\int^{\epsilon}_{-\epsilon}dx\ u(x).
\label{conti}
\end{equation}
However, the integrations in Eq.(\ref{conti}) vanish in the limit 
$\epsilon\to0$ due to 
the continuity of $u(x)$. 
A similar argument applies to $v(x)$ and we have conditions,
\begin{eqnarray}
u'(0+0)-u'(0-0)&=&2\bar{m}u(0),\nonumber\\
v'(0+0)-v'(0-0)&=&-2\bar{m}v(0).\label{eq:BC1}
\end{eqnarray}
We look for solutions under these conditions. 

Let the energy eigenvalue $E$ satisfy
\begin{equation}
\bar{m}-m_0>E>0,
\end{equation}
then $u(x)$ takes the form
\begin{eqnarray}
u(x)&=&\left\{
\begin{array}{ccc}
A\ e^{\kappa_1 x}+B\ e^{-\kappa_1 x}&\ &(x<0),\\
C\ e^{\kappa_2 x}+D\ e^{-\kappa_2 x}&\ &(0<x),
\end{array}
\right.\\
\kappa_1&=&\sqrt{(\bar{m}+m_0)^2-E^2},\\
\kappa_2&=&\sqrt{(\bar{m}-m_0)^2-E^2}.
\end{eqnarray}
From the continuity of $u(x)$,
\begin{equation}
A+B=C+D
\end{equation}
and Eq.(\ref{eq:BC1}) gives
\begin{equation}
\kappa_2(C-D)-\kappa_1(A-B)=2\bar{m}(A+B).
\end{equation}
Hence,
\begin{equation}
u(x)=Ae^{\kappa_1x}+Be^{-\kappa_1x},\hspace{4mm}{\rm for}\hspace{4mm}x<0\\
\end{equation}
and for $x>0$,
\begin{eqnarray}
u(x)&=&\left[\ \frac{2\bar{m}-\kappa_1+\kappa_2}{2\kappa_2}A+\frac{2\bar{m}
+\kappa_1
+\kappa_2}{\kappa_2}B\ \right]\ e^{\kappa_2x}\nonumber\\
&&\nonumber\\
&&+\left[\ \frac{-2\bar{m}+\kappa_1+\kappa_2}{2\kappa_2}A+\frac{-2\bar{m}
-\kappa_1
+\kappa_2}{\kappa_2}B\ \right]\ e^{-\kappa_2x}.
\label{eq:solution4}
\end{eqnarray}
If we let $m_0=0$, $\kappa_1=\kappa_2$ and 
Eq.(\ref{eq:solution4}) simplifies to
\begin{equation}
u(x)=\left\{
\begin{array}{cc}
A\ e^{\kappa x}+B\ e^{-\kappa x}, &\hspace{8mm}(x<0)\\
&\\
\displaystyle{\left[\ \frac{\bar{m}}{\kappa}A+
\frac{\bar{m}+\kappa}{\kappa}B\ \right]\ e^{\kappa x}+
\left[\ \frac{-\bar{m}+\kappa}{\kappa}A-
\frac{\bar{m}}{\kappa}B\ \right]e^{-\kappa x}}. &\hspace{8mm}(0<x)
\end{array}
\right.
\end{equation}
Values of $A,\ B$ and $E$ are determined by boundary conditions, 
e.g., $u(-L/2)=u(L/2)=0$ for a given system size $L$.

Our assumption that $u(x)$ remains continuous in the limit $\lambda\to\infty$ 
is justified as follows. 
The Schr$\ddot{\mbox{o}}$dinger equation with the potential
\begin{eqnarray}
V(x)&=&\pm m'(x)+m^2(x), \nonumber\\
m(x)&=&\bar{m}\tanh(\lambda\bar{m}x)
\end{eqnarray}
is transformed to hyper-geometric differential equation by the change of
variables,
\begin{equation}
z=\frac1{1+e^{-2\lambda\bar{m}x}},
\end{equation}
and solved exactly. If we let $\lambda$ tend to infinity, 
the solution coincides with (\ref{eq:solution4}).
\subsection{Transfer-matrix formalism}
In this subsection we restrict $m_0$ to be zero. 
Since the wave function  $u(x)$ is expressed everywhere as
\begin{equation}
u(x)=A\ e^{\kappa x}+B\ e^{-\kappa x},
\end{equation}
we represent the eigenfunction in terms of coefficients $A$ and $B$ 
in what follows. 
By using matrix representation and  from (\ref{eq:solution4}), the  
conditions (\ref{eq:BC1}) gives the following relation between 
the coefficients $A$ and $B$,
\begin{eqnarray}
\left(
\begin{array}{c}
A(x>0)\\
B(x>0)
\end{array}
\right)&=&T\left(
\begin{array}{c}
A(x<0)\\
B(x<0)
\end{array}
\right), \nonumber\\
&&\nonumber\\
T&=&\left(
\begin{array}{cc}
\displaystyle{1+\frac{\bar{m}}{\kappa}}&\displaystyle{\frac{\bar{m}}{\kappa}}\\
&\\
\displaystyle{-\frac{\bar{m}}{\kappa}}&\displaystyle{1-\frac{\bar{m}}{\kappa}}
\end{array}
\right).\label{eq:defineT}
\end{eqnarray}
For  soliton instead of anti-soliton, one should replace $\bar{m}$ with 
$(-\bar{m})$ in Eq.(\ref{eq:defineT}) and one obtains
\begin{equation}
\left(
\begin{array}{c}
A(x>0)\\
B(x>0)
\end{array}
\right)=T^{-1}\left(
\begin{array}{c}
A(x<0)\\
B(x<0)
\end{array}
\right).
\end{equation}
Let us define
\begin{eqnarray}
R(\kappa,a)&\equiv&\left(
\begin{array}{cc}
e^{\kappa a}&0\\
0&e^{-\kappa a}
\end{array}
\right),\nonumber\\
\phi&\equiv&\frac{\bar{m}}{\kappa},
\end{eqnarray}
and TM for the configuration of  an anti-soliton and a soliton is
given as 
\begin{eqnarray}
\lefteqn{R(b)T^{-1}R(a)T}\nonumber\\
&=&\left(
\begin{array}{cc}
(1-\phi^2)e^{\kappa(a+b)}+\phi^2e^{\kappa(b-a)}&\phi(1-\phi)
\left[e^{\kappa(a+b)}-e^{\kappa(b-a)}\right]\\
\phi(1+\phi)\left[e^{\kappa(a-b)}-e^{-\kappa(a+b)}\right]&(1-\phi^2)
e^{-\kappa(a+b)}+\phi^2e^{\kappa(a-b)}
\end{array}
\right)\nonumber\\
&=&R(b)\left(
\begin{array}{cc}
e^{\kappa a}-2\phi^2\sinh\kappa a&2\phi(1-\phi)\sinh\kappa a\\
2\phi(1+\phi)\sinh\kappa a&e^{-\kappa a}+2\phi^2\sinh\kappa a
\end{array}
\right)\nonumber\\
&\equiv&T(a,b),\label{eq:tmatrtix1}
\end{eqnarray}
where $a$ is the distance between the soliton and anti-soliton.
We impose boundary condition on the wave function such
that {\em the wave function vanishes as $|x|$ tends to large}. 
This means that the solution, which is proportional to $e^{\kappa x}$ 
for $x\to-\infty$,
 does not have the factor of $e^{\kappa x}$ for $x\to\infty$. 
Hence $\kappa$ needs to satisfy
\begin{equation}
\left(
\begin{array}{cc}
1&0
\end{array}
\right)T(a,b)\left(
\begin{array}{c}
1\\0
\end{array}
\right)=e^{\kappa b}(e^{\kappa a}-2\phi^2\sinh\kappa a)=0.
\end{equation}
Since $e^{\kappa b}\ne0$, this means
\begin{equation}
\frac{\kappa^2}{\bar{m}^2}=1-e^{-2\kappa a}.
\end{equation}
The energy eigenvalue $E$ tends to vanish in the limit $\kappa\to \bar{m}$, 
$ a \to \infty$, as expected. 
If there are two pairs of  anti-soliton and soliton, the boundary 
condition leads to
\begin{equation}
\left(
\begin{array}{cc}
1&0
\end{array}
\right)T(c,d)T(a,b)\left(
\begin{array}{c}
1\\0
\end{array}
\right)=0,
\end{equation}
or
\begin{eqnarray}
0&=&e^{\kappa(a+b+c)}+2\phi^2[\ \sinh\kappa a\ \sinh\kappa 
c\ e^{-\kappa b}-e^{\kappa(b+c)}\sinh\kappa a-e^{\kappa(a+b)}
\sinh\kappa c\ ]\nonumber\\
&&+8\phi^4\sinh\kappa a\sinh\kappa b\sinh\kappa c.
\end{eqnarray}
The above argument can be generalized to an arbitrary numbers of 
anti-soliton-soliton 
pairs readily and we have the following eigenvalue equation, 
\begin{equation}
\left(
\begin{array}{cc}
1&0
\end{array}
\right)T(e,f)\cdots T(c,d)T(a,b)\left(
\begin{array}{c}
1\\0
\end{array}
\right)=0. \label{eq:tmatrtix2}
\end{equation}
It is easily seen that variables $\kappa$, $\bar{m}$ and distances 
between solitons $a,b,c,\cdots$ appear always in the combinations 
$\bar{m}/\kappa$, $\kappa a$,{\it etc.} 
Hence the eigenvalue equation (\ref{eq:tmatrtix2}) 
is invariant under a transformation
\begin{equation}
E\to\alpha E,\hspace{4mm}\bar{m}\to\alpha\bar{m},\hspace{4mm}a\to 
a/\alpha\label{eq:invariance1}
\end{equation}
with an arbitrary positive constant $\alpha$. 
This property becomes important when we discuss relation between the 
localization 
length and the energy dispersion as we shall see in later sections. 
We are not able to solve (\ref{eq:tmatrtix2}) analytically for arbitrary 
configuration of pairs of solitons. 
However it is not so difficult to solve (\ref{eq:tmatrtix2}) numerically.
We can also easily obtain eigenfunctions after having eigenvalues
by using the TM.
In subsequent sections, we shall give solutions to Eq.(\ref{eq:tmatrtix2}) 
obtained 
numerically for various multi-soliton configurations $m(x)$.
Then we shall compare the results with the analytical calculation by the 
SUSY methods
in our previous paper\cite{IK}.
It is straightforward to extend the above formalism for nonvanishing $m_0$.
Numerical solutions will be given also for $m_0\neq 0$.

\setcounter{equation}{0}
\section{Results of numerical calculation}

\subsection{Overview}
We solved (\ref{eq:tmatrtix2}) numerically in various multi-soliton 
backgrounds, and obtained explicit form of corresponding eigenfunctions.
We are interested in the relationship between randomness
and localization of eigenfunctions.
Then we compare wave functions of
the eigenstates in quasi-periodical backgrounds and random backgrounds.

\begin{enumerate}
  \item Quasi-Periodic Background

  First, we consider a quasi-periodic background in a system of 
  finite size $L$, with 
  almost equal distances between each successive soliton and anti-soliton. 
  Eigenfunctions obtained by numerical calculations show 
  twofold oscillations, {\it i.e.}, rapid oscillations occurring at 
  positions of soliton and anti-soliton, 
  and slowly oscillating envelope (see the graphs at the top of Fig.2). 
  The peaks of rapid oscillation can be considered essentially the same as 
  the bound state (\ref{eq:boundstate}) in a single soliton background, 
  because in a background of 
  sufficiently separated solitons, eigenfunctions can be approximated 
  by a linear combination of bound states located at solitons and 
  anti-solitons.

 The envelope of eigenfunctions is reminiscent of sine curves and the nodes 
 of 
 eigenfunctions coincide with those of corresponding sine curves. 
 Frequencies of 
 the envelopes are larger for higher energy eigenvalue and this supports
  ``one-dimensional node counting theorem"\cite{Comtet}, 
  which states that the number of nodes of eigenfunction increases with  
  eigenenergy.

In quasi-periodic background, all eigenstates extend over the whole system 
and there are no localized eigenstates. 
This is reminiscent of Bloch's theorem  
although we do {\em not} impose the periodic boundary 
condition on the wave functions.

\begin{figure}
\label{fig:quasiperiod1}
\begin{center}
\unitlength=1cm
\begin{picture}(15,15)
\unitlength=1mm
\centerline{
\epsfysize=15cm
\epsfbox{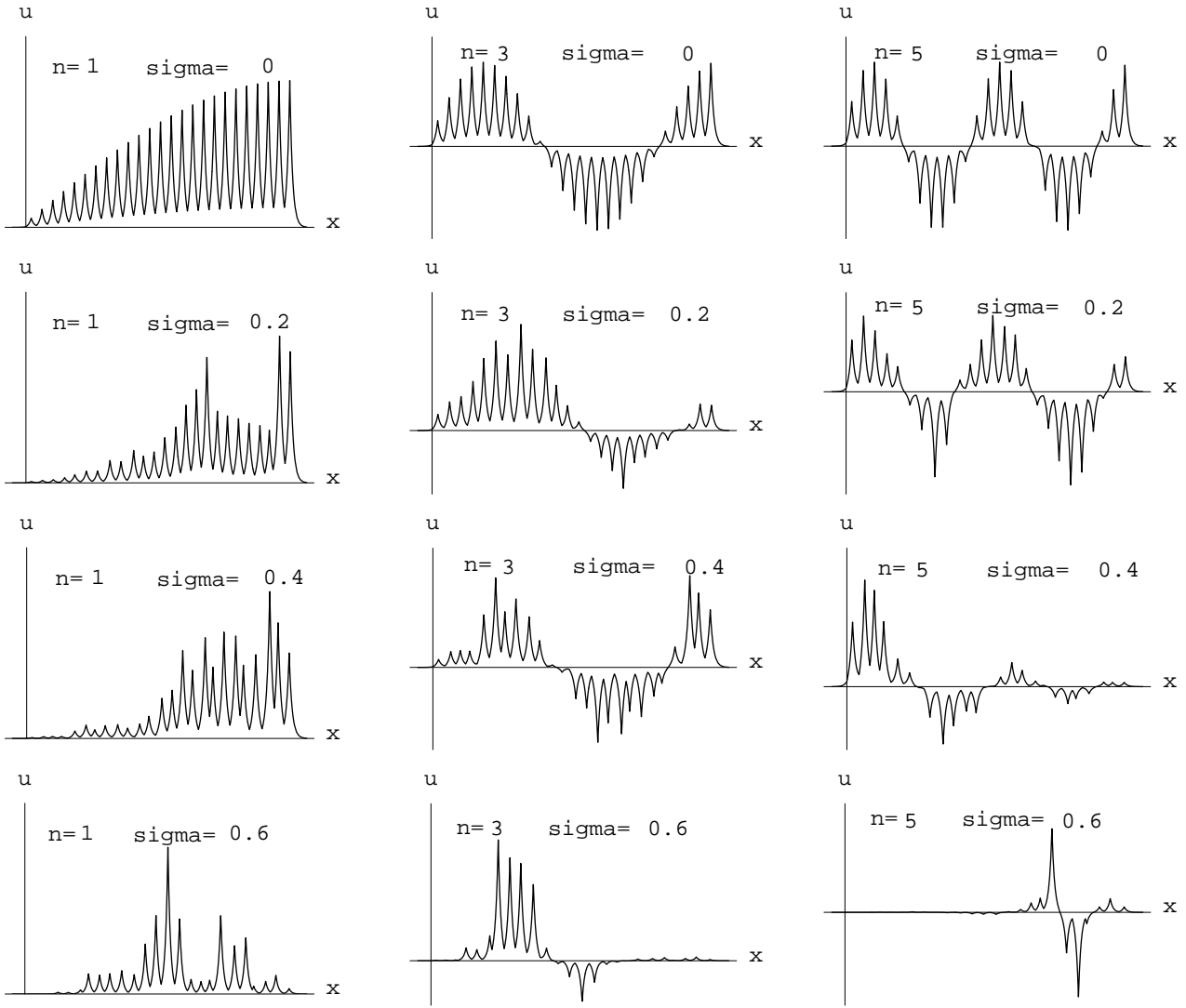}
}
\end{picture}
\caption{The shapes of $u(x)$ in rectangular barriers distributed
according to gaussian 
distribution with the average lengths $\langle l\rangle=2$ and $\bar{m}=1$. 
In this figure $n$ denotes the state of $n$-th energy from the lowest and
 $m_0$ is set to zero. 
The graphs at the top are those in a periodic potential, where the standard 
deviation $\sigma=0$. Here $\sigma$ increases from zero at the top to $0.6$ 
at the bottom by $0.2$.
Eigenfunctions $v(x)$ have similar behavours to $u(x)$.}
\end{center}
\end{figure}
 
  \item Randomly Distributed Barriers: $m_0=0$

We now vary the distances  between soliton and anti-soliton according to the
gaussian distribution, {\it i.e.}, lengths of rectangular barriers 
$l$ are subject to the gaussian distribution,
\begin{equation}
P_{{\rm G}}(l)=\frac1{\sqrt{2\pi}\sigma}\exp\{-(l-\mu)^2/2\sigma^2\}.
\label{gdist}
\end{equation}
$m_0$ is set to zero here. 
In this case, the periodicity of eigenfunction is lost as randomness or 
$\sigma$ 
in Eq.(\ref{gdist}) increases and the envelope has a large peak.
This means all the states tend to ``localize" (see Fig.2). 
The state of the lowest energy $E\simeq0$ extends over the whole system 
as it can be expected from the formula (\ref{eq:exactsol}). 
Low-energy states are still extended even at $\sigma=0.4$.

\begin{figure}
\label{fig:exponential1}
\begin{center}
\unitlength=1cm
\begin{picture}(15,10)
\centerline{
\epsfysize=10cm
\epsfbox{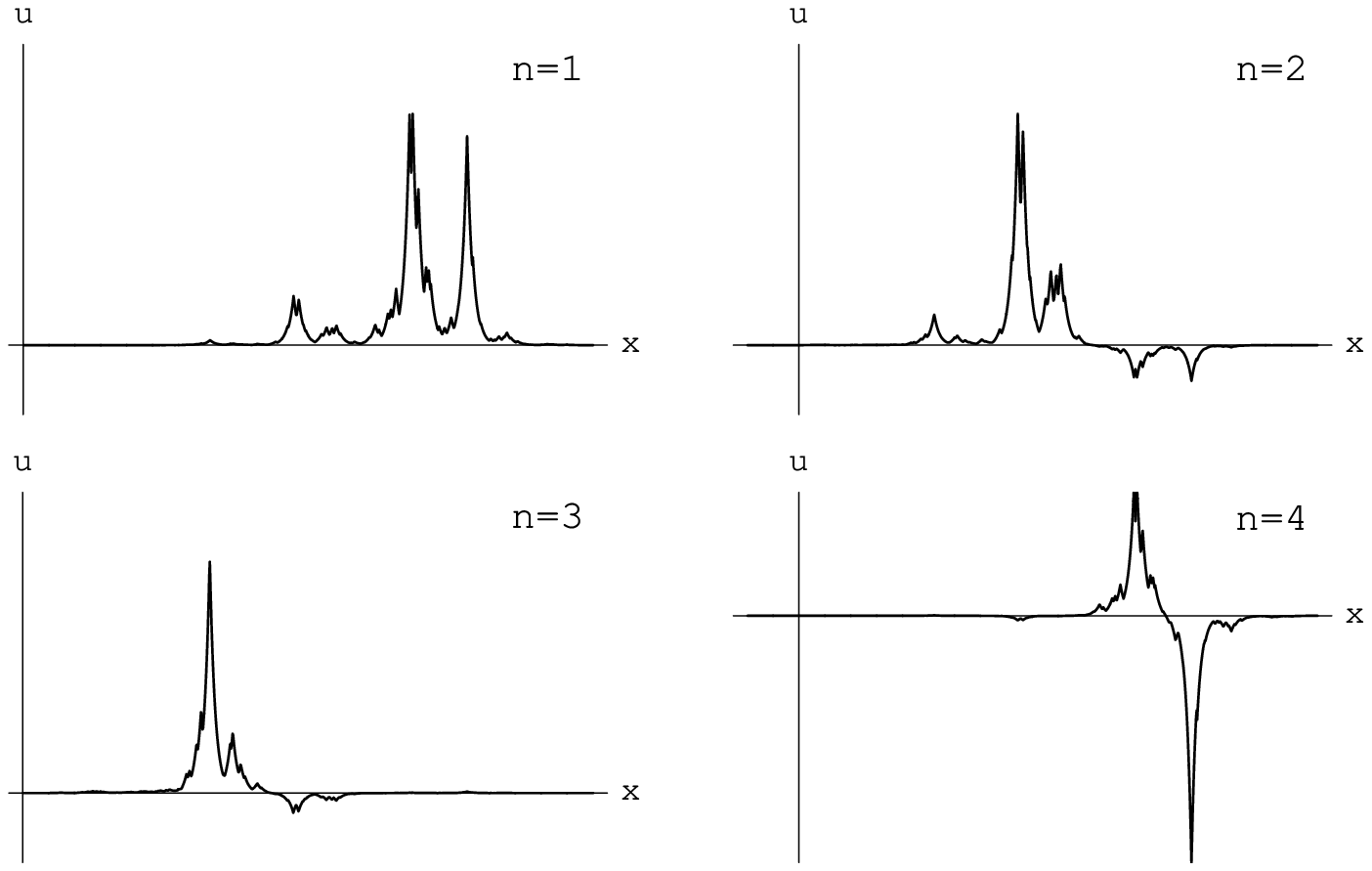}
}
\end{picture}
\caption{The shapes of $u(x)$ in rectangular barriers 
whose lengths $l$ are distributed according to 
exponential distribution with $\tilde{\lambda}=0.25$ 
(Eq.(\ref{eq:exponentialdist})). 
In this figure $n$ denotes the state of $n$-th energy from the lowest. 
$\bar{m}$ and $m_0$ are set as $m_0=0$ and $\bar{m}=1$.}
\end{center}
\end{figure}
  The degree of randomness is controlled by the variance of 
  gaussian distribution 
  $\sigma$. 
  For larger $\sigma$, shapes of wavefunctions become sharper and the 
  states become more localized. 
  We took a close look at various numerical results
  and found that some of the 
  states have more than one dominant peak in two or three intervals 
  between nodes and others have only single ``large"  peak and
  are localized in an interval between adjacent nodes. 
  Let $N(E)$ denotes the number of states from the lowest 
  to the energy $E$ per unit length. 
  Since the width of the interval is approximately proportional to $1/N(E)$,
  we are led to a conjecture that the localization length 
  $\xi(E)$ satisfies\footnote{As is well known, there are two 
  different localization lengths in the present system, i.e.,
  the typical and disorder-averaged localization lengths 
  (Griffiths phase).
  In this discussion, $\xi(E)$ is considerded as the averaged 
  localization length, which is defined by the long-distance behaviour
  of disorder-averaged Green's functions, i.e.,
  we expect that the Green's functions are dominated by 
  the states with more than one peak\cite{BF}.}
\begin{equation}
\xi(E)=\frac{A}{N(E)}.
\end{equation} 
The proportional coefficient $A$ seems to depend on the distribution of 
the length 
of rectangular barriers $l$, $\bar{m}$ and $L$, but not on $E$.

In Fig.3, we show eigenfunctions in random rectangular barriers 
whose lengths $l$ 
are distributed according to the following exponential distribution 
instead of (\ref{gdist}),
\begin{equation}
P_{{\rm E}}(l)=\frac1{2\tilde{\lambda}}\exp\left[-\frac{l}{2\tilde{\lambda}}
\right].\label{eq:exponentialdist}
\end{equation}
From these results, we can see that the localization around one 
peak is enhanced
in higher-energy states.
We also verified that as $\tilde{\lambda}$ increases extended 
states are hindered,
just as in the gaussian distribution.

\item For $m_0\ne0$ 

 In the above we dealt only with the case of vanishing $m_{0}$. 
 Here, we take $m_0\neq 0$.
For quasi-periodic backgrounds, states are still periodic and extended.
``Bloch's theorem" is valid as  in the case of $m_0=0$. 
If we vary the length of barrier $l$ randomly,
states are more localized than in the case of  vanishing $m_0$ (see Fig.4). 
Thus we conclude that nonzero $m_{0}$ is not necessary for localization, 
but at least it seems to enhance the localization.
\begin{figure}
\label{fig:nonzerom1}
\begin{center}
\unitlength=1cm
\begin{picture}(15,10)
\centerline{
\epsfysize=10cm
\epsfbox{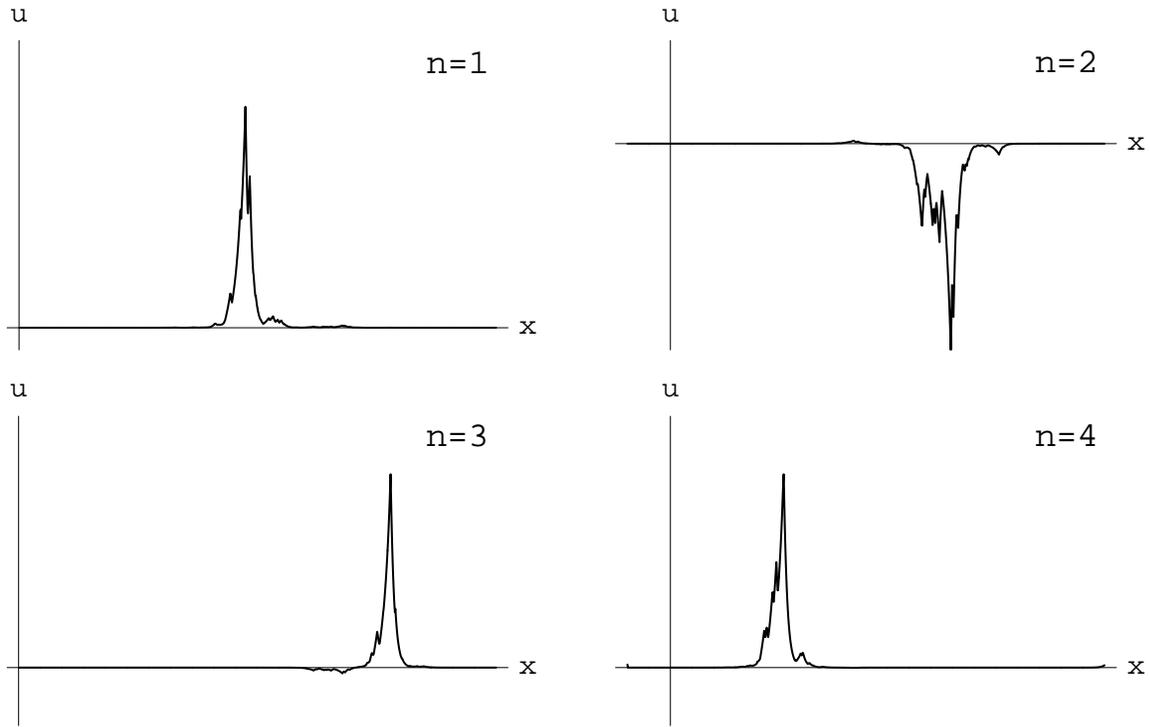}
}
\end{picture}
\caption{The shapes of $u(x)$ in rectangular barriers 
whose lengths $l$ are distributed according 
to exponential distribution with $\tilde{\lambda}=0.25$ 
(Eq.(\ref{eq:exponentialdist})). 
In this figure $n$ denotes the state of $n$-th energy from the lowest. 
The parameters $\bar{m}$ and $m_0$ are set as $m_0=0.3$ and $\bar{m}=1$.}
\end{center}
\end{figure}
\end{enumerate}


\subsection{Comparison with the analytical results: Density of states}

 Although Dirac fermion systems with non-locally correlated mass $m(x)$ have 
 seldom been studied, 
 the case of exponentially distributed $m(x)$ was studied recently 
 by Comtet et.al.\cite{Comtet} via stochastic method and more 
 systematically by 
 Ichinose and Kimura \cite{IK} via SUSY methods. 
 Analytical expression of the DOS was obtained there. 
 We calculate the DOS numerically by using the TM method 
 and compare the result with analytical expression.

For {\em locally} distributed $m(x)$, the density of states with energy 
$E$, $\rho(E)={dN(E) \over dE}$,
 is given by 
 \begin{equation}
 \label{eq:spectrum1}
  \rho(E) \propto \frac{1}{|\ln E|^3},
 \end{equation}  
as $E$ tends to small. 
For {\em non-locally} correlated random mass $m(x)$\cite{IK},
 \begin{eqnarray}
 \label{eq:spectrum2}
  \rho_{\tilde{\lambda}}(E)&=&\hspace{4mm}
\frac{1}{2\frac{E}{4g}
 |\ln \frac{E}{4g}|^3} 
  \nonumber \\
&&  
-4g\tilde{\lambda} \frac{3}{2\frac{E}{4g}|\ln \frac{E}{4g}|^4}
  \nonumber \\
&&+16g^2\tilde{\lambda}^2\left(\frac{13}{30\frac{E}{4g}|\ln\frac{E}{4g}|^3}
-\frac{1}{2\frac{E}{4g}|\ln\frac{E}{4g}|^4}
+\frac{1}{\frac{E}{4g}|\ln\frac{E}{4g}|^5}\right) 
 \nonumber \\
&&-(4g)^3\tilde{\lambda}^3\frac1{210\frac{E}{4g}|\ln\frac{E}{4g}|^3}+ 
{\cal O}(\tilde{\lambda}^4),
 \end{eqnarray}
where $g$ and $\tilde{\lambda}$ are defined as\footnote{Definitions 
of $g$ in Refs.\cite{BF} and \cite{IK} are in fact different by
factor $2$. In this paper we follow that in \cite{BF}.}
 \begin{equation}
 [\ m(x)\ m(y)\ ]_{{\rm ens}} = \frac{g}{\tilde{\lambda}} 
 \exp\ (-|x-y| / \tilde{\lambda}).
 \label{disordercor}
 \end{equation} 
 Parameter $\tilde{\lambda}$ is the correlation length of the disorder. 
 As $\tilde{\lambda} \rightarrow
 0$, $m(x)$ and $m(y)$ for $x\neq y$ become uncorrelated, 
 the white-noise limit.

Distribution corresponding to (\ref{disordercor}) can be realized by 
rectangular barriers 
whose height is
\begin{equation}
\bar{m}=\sqrt{\frac{g}{\tilde{\lambda}}},
\end{equation}
and  width between solitons, $l$, is distributed as\cite{Comtet}
\begin{equation}
P(l)=\frac1{2\tilde{\lambda}}\exp\left[-\frac{l}{2\tilde{\lambda}}\right].
\end{equation}

The result of numerical calculation of 300 pairs of a soliton and 
anti-soliton is shown in Fig.5. 
We obtained the number of states $N_{num}(E)$ by 
numerical calculation, 
and we define $\rho_{num}$ as the averaged density of states
around the energy level $E$ as
\begin{equation}
\rho_{num}(E) = \frac{ N_{num}(E+\frac{\Delta E}{2})-
  N_{num}(E-\frac{\Delta E}{2}) } { \Delta E }.
\end{equation}
In order to compare the numerical result of the density of states 
with energy $E$ with the 
analytical expression $\rho_{\tilde{\lambda}} (E)$ in (\ref{eq:spectrum2}),
 we see the ratio $ r(E) = \frac {\rho_{\tilde{\lambda}}(E)}
{\rho_{num}(E)} $. 
If the energy dependence of these density of states coincides,
this ratio $r(E)$ should be constant.
In Fig.5. $r(E)$ is shown. 
From this, we can conclude that
the energy spectrum of the states obtained by
the numerical calculation is in good agreement with the analytical result 
 (\ref{eq:spectrum2}). 
Especially, the higher-order expression of  $\rho_{\tilde{\lambda}}(E)$ 
in (\ref{eq:spectrum2}) gives the better agreement. 
This result also indicates that 300 pairs of a soliton and anti-soliton 
is large enough for the investigation of the system in the 
low-energy region.

\begin{figure}
\begin{center}
\unitlength=1cm
\begin{picture}(15,10)
\centerline{
\epsfysize=10cm
\epsfbox{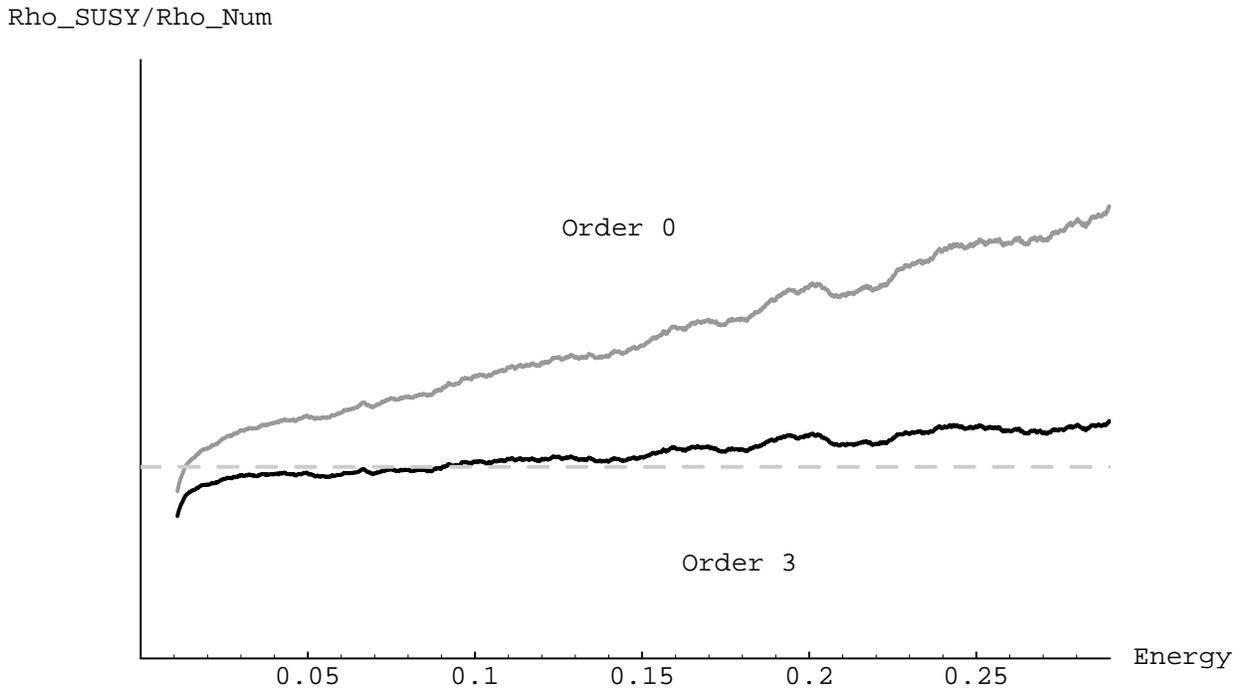}
}
\end{picture}
\caption{Energy spectrum of fermion: 
A comparison between the result of numerical calculation and 
$\rho_{\tilde{\lambda}}(E)$
of the zero-th order and up to the third order in $\tilde{\lambda}$, 
Eq.(\ref{eq:spectrum2}). 
Here we set $\tilde{\lambda}=1/6$, $g=1$,
$\Delta E = 0.02$
and numerical result is averaged over 4000 trials.
The ratio $r(E)$ is shown in this figure.
If the numerical calculation agrees with $\rho_{\tilde{\lambda}}(E)$
which is obtained by SUSY methods, the ratio should be constant. 
The dashed line is expected exact result.}
\end{center}
\label{fig:ratio}
\end{figure}


\subsection{Green's functions at vanishing energy}

In this section, we show numerical result of the ensemble
averaged Green's functions at vanishing energy and compare them
with analytical expression in Refs.\cite{BF,Shelton}.

The zero-energy wave functions are given by Eq.(\ref{eq:exactsol}).
It is not so difficult to calculate the ensemble-averaged
correlation of them 
\begin{equation}
W_q(x,L)=[|\psi^{\dagger}(x)\psi(0)|^q]_{{\rm ens}}
\label{Wq}
\end{equation}
in the white-noise limit with the system size $L$\cite{BF,Shelton}.
There the normalization of the wave functions plays an 
important role because only for specific $m(x)$ they are
genuine normalizable functions.
More explicitly, in solving the Dirac equation  
we imposed the boundary condition such that the wave function
decays exponentially outside of the random potential.
In the analytical calculation in \cite{BF,Shelton}, such kind of
boundary condition is not imposed on the wave function.   

In general it is expected that the above correlation functions
exhibit complex multi-fractal scaling as in the qunatum Hall state,
\begin{equation}
W_q(x,L) \sim L^{-d-\tau(q)}|x|^{-y(q)}.
\label{multiF}
\end{equation}
For the present system, $W_q(x,L)$ is obtained as follows 
for $L \rightarrow \infty$,
\begin{equation}
W_q(x,L) \sim \frac{\tilde{W}(q^2x)}{L}
\label{Wq2}
\end{equation}
with
\begin{equation}
\tilde{W}(x)=\int^{\infty}_0dk \frac{k^2}{(1+k^2)^4}e^{-xk^2}.
\label{tildeW}
\end{equation} 
For large $x$ 
\begin{equation}
W_q(x,L) \sim 1/(x^{3/2}L)
\end{equation}
and therefore it is expected that $\tau(q)=0$ and $y(q)=3/2$.

We numerically calculated the correlation functions $W_q(x,L)$
with various values of $q$. 
To this end, we focused on the states of nodeless wave function.
The results are shown in Figs.6
and 7 for $q=0.5,\; 1.0,\; 1.2\;$ and $1.5$
as well as the analytical result Eqs.(\ref{Wq2}) and 
(\ref{tildeW}).
It is obvious that the numerical calculations are in good agreement
with the analytical calculations.
Especially we can conclude that the critical exponent of the 
multi-fractal scaling is obtained as $y(q)=3/2$ independently of $q$.
(It seems that $y(q)$ is smaller than $3/2$ for small $q$.
 But we expect that $y(q)$ with any $q$ approaches to $3/2$
 for larger systems.) 

We also study $L$ dependence of $W_{q}(x,L)$. The result is shown in
Fig.8. According to this, it seems that $W_{q}(x,L)$
is not inversely proportional to $L$ and moreover $\tau(q)\neq 0$,
contrary to the analytical calculation by Balents and Fisher\cite{BF}.
We further calculated $W_{q}(x,L)$ for larger system size $L$ and 
smaller $g$,
and obtained similar results.
However it is still possible that the system size $L$ is not large enough
in our numerical calculations.
Therefore the above is {\em not}
a definite conclusion and more systematic calculation is desired.
 (In the analytical calculation\cite{BF},
 $L$ is assumed to be large enough.) 

We should again remark here that Eqs.(\ref{Wq2}) and (\ref{tildeW})
are obtained from (\ref{eq:exactsol}) some of which are {\em not}
normalizable.
On the contrary, the wave functions in the numerical calculation
are all normalizable.
Then strictly speaking, one cannot expect coincidence of the
analytical and numerical results.

\begin{figure}
\label{fig:fractal1}
\begin{center}
\unitlength=1cm
\begin{picture}(15,15)
\centerline{
\epsfysize=15cm
\epsfbox{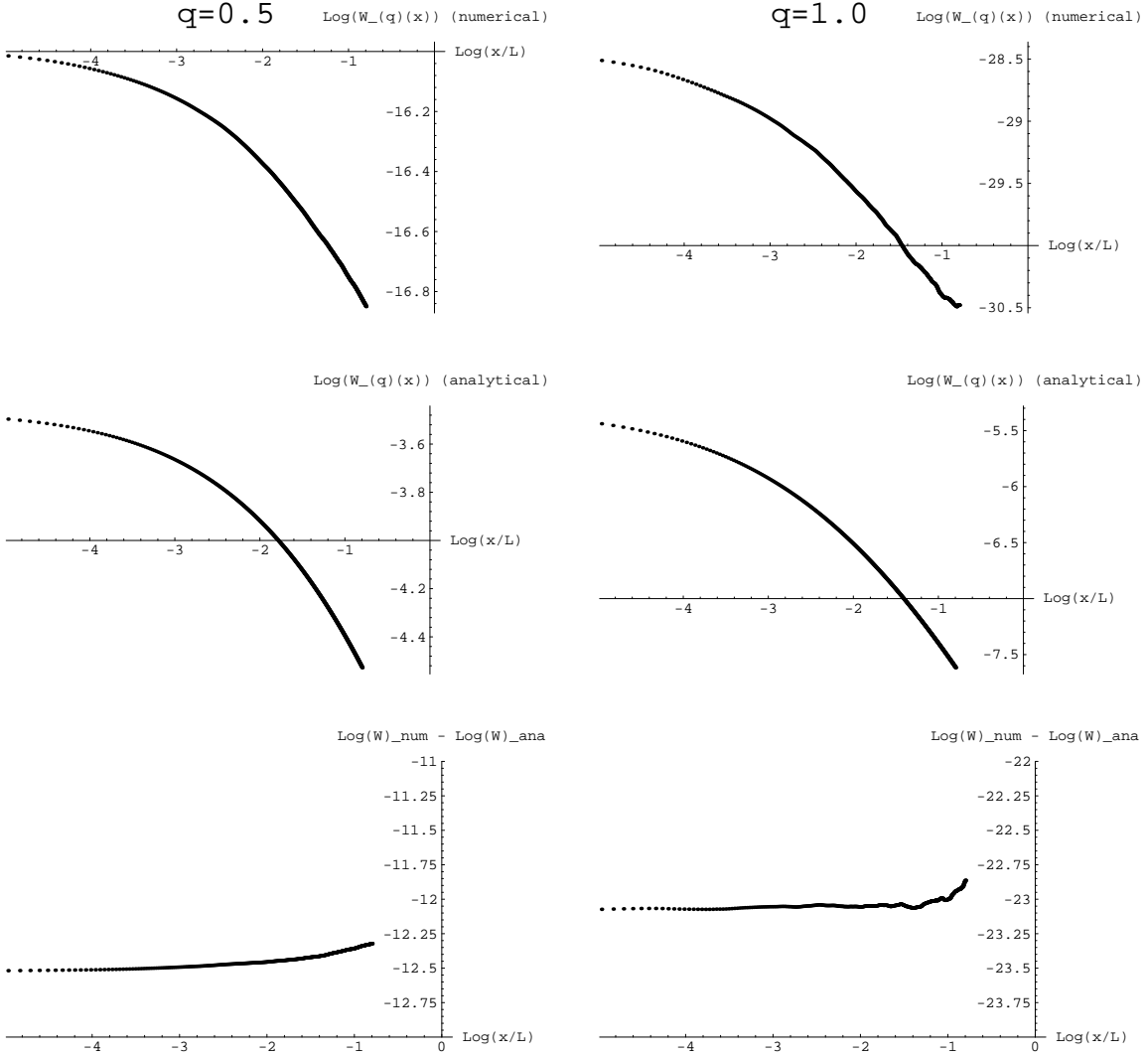}
}
\end{picture}
\caption{Multifractal behaviour of bound-state wavefunction (1) :
The $x$ dependence of 
$W_{q}(x,L)$ based on numerical and analytical 
calculations is shown.  
Difference between two results is also shown. 
To calculate $W_{q}(x,L)$, we used the u-component of
the doublet $\psi=(u,v)$, namely  $W_{q}(x,L)_{num}=
[|u(x+l)u(l)|]_{\rm ens}$. 
To obtain $W_{q}(x,L)_{num}$ numerically, 
we averaged $W_{q}$ over 2800 configurations of the random-varying mass.
We also took an average with respect to
 $l$ in $[|u(x+l)u(l)|]_{\rm ens}$.}
\end{center}
\end{figure}

\begin{figure}
\label{fig:fractal2}
\begin{center}
\unitlength=1cm
\begin{picture}(15,15)
\centerline{
\epsfysize=15cm
\epsfbox{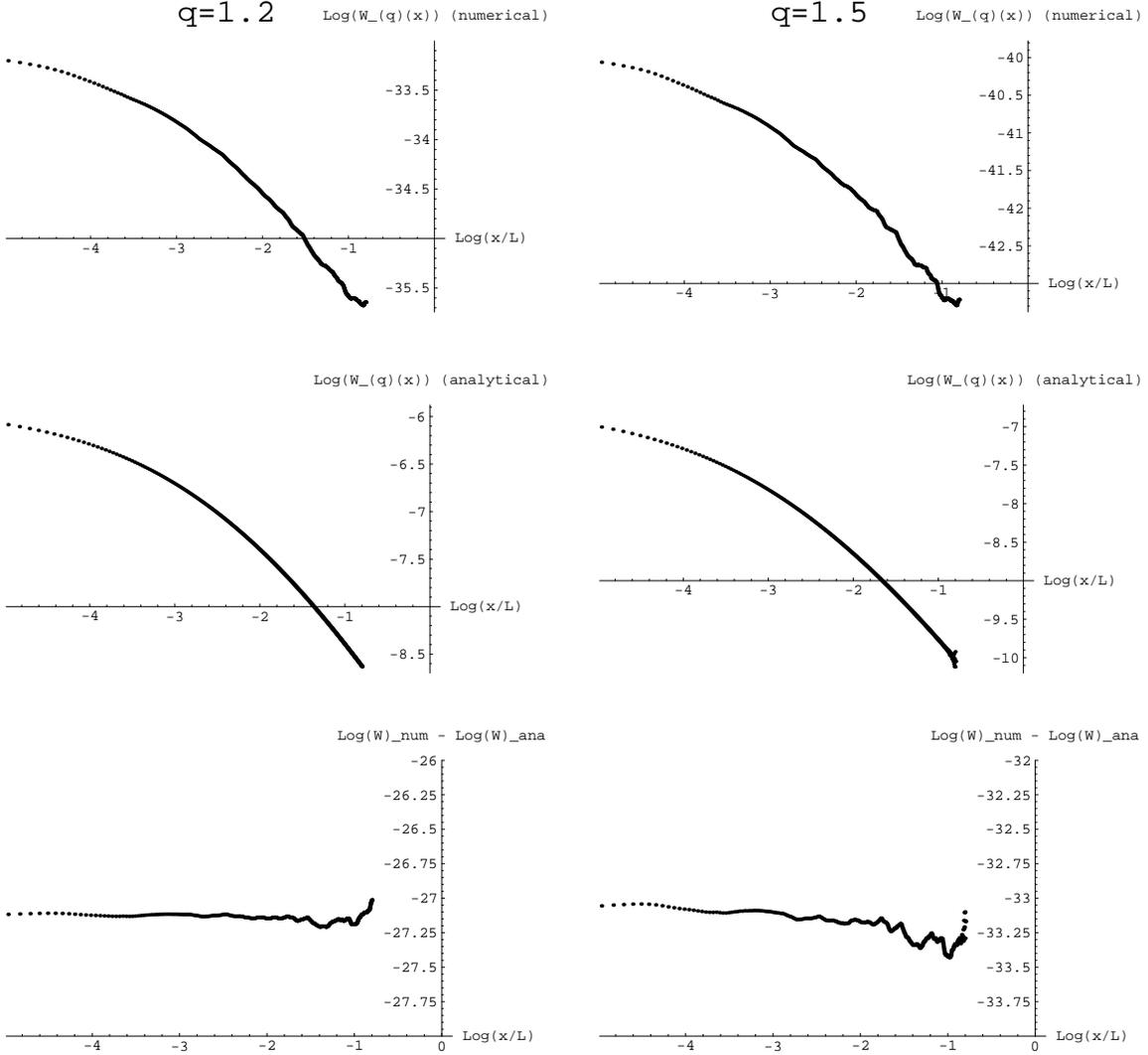}
}
\end{picture}
\caption{Multifractal behaviour of bound-state wavefunction (2) :
The method to obtain $W_{q}$ is the same as we mentioned
in the caption of Fig.6.
We can see that the inclination of $W_{q}$ with $q=1.2$ and $1.5$
is almost 3/2 in the region around $\log(x/L) \sim -1$.  
(We set $L$(the system size)$=40$,
$\tilde{\lambda}=1/512$ and $g=1/2$ (almost 
white-noise case) in the numerical calculation,
and we set $L=40$ and $g=1/2$ in the analytical calculation.)
}
\end{center}
\end{figure}

\begin{figure}
\label{fig:fractal3}
\begin{center}
\unitlength=1cm
\begin{picture}(12,15)
\centerline{
\epsfysize=15cm
\epsfbox{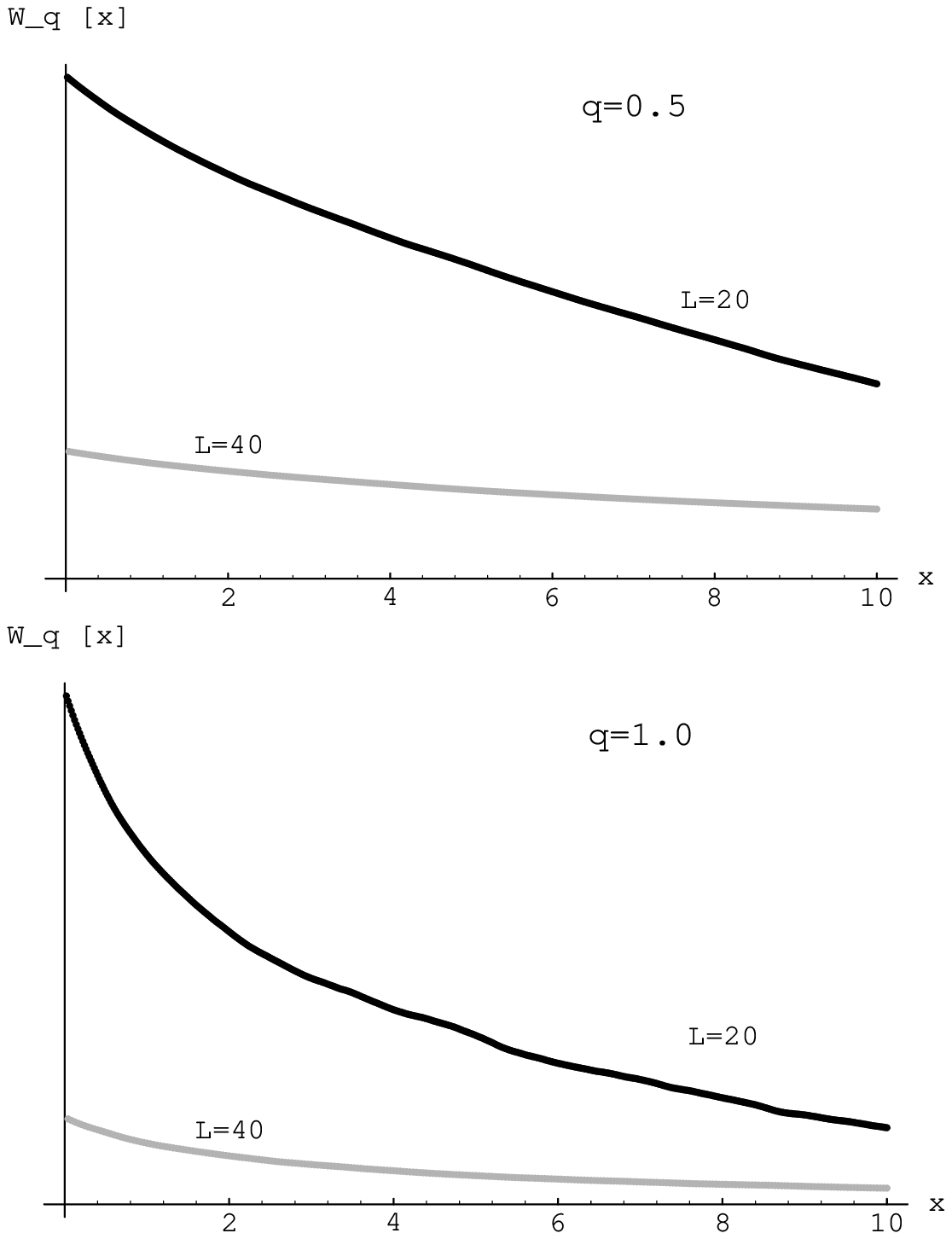}
}
\end{picture}
\caption{Multifractal behaviour of bound-state wavefunction (3) :
$L$(the system size) dependence of $W_{q}(x,L)$ for 
$q=0.5$ and $1.0$ is shown. 
Here we set $\tilde{\lambda}=1/512$ and $g=1/2$, 
and $W_{q}$ is averaged over 1600 backgrounds of the random mass. To
calculate $W_{q}\;$, we used only the u component, though we verfied
that the v component behaves similarly.
} 
\end{center}
\end{figure}

\setcounter{equation}{0}  
\section{Discussion}

From the results of the DOS and the Green's functions, 
we conclude that 300 pairs of a soliton and anti-soliton 
 can be regarded as a good approximation for infinite-size system.
 Hence it is legitimate to discuss behaviors of the localized states 
 based on properties of numerically obtained solutions.

Results obtained from the numerical calculations can be summarized 
as follows.
\begin{itemize}
\item For a quasi-periodic background, ``Bloch's theorem" is valid and 
all the states extend over the whole system, regardless of the value 
of $m_0$. 
The envelopes of wave functions are similar to sine curves.
\item For $m_0=0$, on varying the length of rectangular barriers randomly, 
states begin to localize, within two or three intervals between nodes.
Some of them, especially the lowest-energy states,
has more than one peak which separate with each other rather
long distance.
\item For nonzero value of $m_0$, the states tend to localize in narrower 
intervals 
than in the case of $m_0=0$. 
Hence nonzero $m_0$ enhances, but is not necessary for the localization. 
\end{itemize}

We conjectured that the averaged localization length $\xi(E_n)$ 
for the $n$-th state from the lowest satisfies 
\begin{equation}
\xi(E_n)=\frac{A}{n},
\label{eq:conjecture1}
\label{EN}
\end{equation}
since states in randomly distributed rectangular barriers are localized in 
two or three intervals between nodes and length of each interval 
$l(E_N)$ is proportional to $1/N$. 
Here, $A$ in (\ref{EN}) depends on the distribution of lengths of 
rectangular barriers $l$, 
$\bar{m}$ and $L$, but not on energy $E$. 
One of supports for this conjecture is given by the exact result for locally 
correlated random mass 
$m(x)$, since the number of states below energy $E$ is given as
\begin{equation}
N(E)\propto\frac1{(\ln E)^2},
\end{equation}
and localization length is given as
\begin{equation}
\xi(E)\propto(\ln E)^2,
\end{equation}
hence (\ref{eq:conjecture1}) is satisfied in this case.
(Note that $N(E)$ and $n$ are related as $N=n/L$.)
Very recently, we obtained the localization length of the
ensemble-averaged one-particle Green's functions by using the SUSY
methods\cite{IK2},
\begin{equation}
\xi(E)={1 \over g\pi^2}\Big(|\ln {E \over 2g}|^2
+4g\tilde{\lambda}|\ln {E \over 2g}|\Big)+O((g\tilde{\lambda})^2).
\label{LL}
\end{equation}
From this result and the DOS in Eq.(\ref{eq:spectrum2}),
we can verify 
\begin{equation}
\xi(E)N_{\tilde{\lambda}}(E) \propto 1+
O((g\tilde{\lambda})^2).
\end{equation}

We also performed numerical calculations with the TM method under Dirichlet 
boundary condition, which is $u(0)=u(L)=0$.
However qualitatively features of the localization are the same. 
Another important insight into the Dirac fermion system with random mass 
$m(x)$ obtained 
by the present study is that the $n$-th state from the lowest is related to 
the 
lowest-energy state in system of size $1/N (=L/n)$, since a state localized
inside an interval 
of length $1/N$ must be realized as the ground state of the system of 
length $1/N$. 

The ensemble-averaged Green's functions were also calculated.
They are in good agreement with the analytical results.
Especially, though each wave function exhibits rapid oscillations,
the ensemble-averaged Green's functions have smooth behaviour.

Finally let us comment on the spin systems closely related to the 
random-mass Dirac fermions. As discussed in Refs.\cite{Shelton,Gog}
low-energy excitations in doped spin-ladder or spin-Peierls systems are 
described by the random-mass Dirac fermions. Undoped cases of the 
above systems have an energy gap, and by doping there appear 
midgap states\cite{FM}. Results obtained in this paper suggest that 
properties of the midgap states strongly depend on the randomness
of the doping, {\it i.e.}, almost all midgap states are extended 
for quasi-periodic doping, whereas states tend to localize for 
random doping. In terms of spin-Peierls system (like the compound
$\mbox{Cu}_{1-x}\mbox{Zn}_{x}\mbox{GeO}_3$ \cite{Hase}), 
the localization length $\xi(E)$ is related with the correlation
length of spins, since the existence or absence of our fermion 
is related to the up or down of the spins via Jordan-Wigner 
transformation\cite{SML}. Using our results one will be able to find that  
randomly doped spin-Peierls compounds have the short correlation 
length of spin. 
It is very interesting to see the above behavior by experiments
in which the distribution of impurities is controled. We are now 
calculating Green's functions in having more definte prediction
for the above spin systems. The result will be reported in a 
future publication\cite{IK2}.


\end{document}